\documentclass{appolb}

\usepackage{graphicx}
\usepackage{subfigure}
\usepackage{cite}
\usepackage{epsfig}
\usepackage{amssymb}
\usepackage{amsmath,amsfonts}
\usepackage{stackrel}

\def\MeV{\,\mathrm{MeV}}

\def\fm{\,\mathrm{fm}}

\def\QCD{\,\mathrm{QCD}}
\def\HRG{\,\mathrm{HRG}}

\newcommand{\ignore}[1]{}

\begin{document}
\title{Thermal correlators in the hadron resonance gas: a dual
  Hagedorn distance\thanks{Talk by E. Meg\'{\i}as at ``Excited QCD
    2018'', Kopaonik, Serbia, March 11-15, 2018.}  \thanks{Work
    supported by the Spanish MINEICO and European FEDER funds (grants
    FIS2014-59386-P, FIS2017-85053-C2-1-P and FPA2015-64041-C2-1-P),
    Junta de Andaluc\'{\i}a (grant FQM-225) and Basque Government
    (grant IT979-16). The research of E.M. is also supported by the
    Ram\'on y Cajal Program of the Spanish MINEICO, and by the
    Universidad del Pa\'{\i}s Vasco UPV/EHU, Bilbao, Spain, as a
    Visiting Professor.}
}
\author{E. Meg\'{\i}as$^{\, a,b}$, E.~Ruiz Arriola$^{\, a}$, L.L.~Salcedo$^{\, a}$
\address{$^{a}$Departamento de F\'{\i}sica At\'omica, Molecular y Nuclear and Instituto Carlos I de F\'{\i}sica Te\'orica y Computacional, Universidad de Granada, \\ Avenida de Fuente Nueva s/n,  18071 Granada, Spain }
\address{$^{b}$Departamento de F\'\i sica Te\'orica, 
Universidad del Pa\'is Vasco UPV/EHU, Apartado 644,  48080 Bilbao, Spain}
\\
}
\maketitle
\begin{abstract}
Fluctuations and correlations of conserved quantities in the confined
phase of QCD are a viable way to characterize the existence of exotic
and missing states with given quantum numbers in the hadronic
spectrum. We study a realization of the Hadron Resonance Gas model in
the light quark ({\it uds}) flavor sector of QCD to study the
fluctuations and static correlators of electric charge, baryon number
and strangeness. It is also conjectured an interesting duality between
the correlators at zero temperature, and the fluctuations of
integrated quantities at low temperatures, leading to the appearance
of a dual Hagedorn distance for the former.

\end{abstract}
\PACS{11.10.Wx, 12.38.-t, 12.38.Lg}

\section{Introduction}

A fundamental quantity for the study of the thermodynamics of QCD is
the partition function, which writes
\begin{equation}
Z_{\QCD} = \Tr \, e^{-H_{\textrm QCD}/T}= \sum_n e^{-E_n/T} \,,
\end{equation}
where $E_n$ are the eigenvalues of the QCD Hamiltonian. This
illustrates the connection between the thermodynamics and the spectrum
of QCD. An explicit realization in the confined phase is given by the
Hadron Resonance Gas (HRG) model~\cite{Hagedorn:1984hz} where the
hadronic states compiled by the PDG~\cite{Patrignani:2016xqp} are
considered as stable, non-interacting and point-like
particles. Microscopic models have also been used to compute the meson
and baryon spectra, such as e.g. the Relativized Quark Model
(RQM)~\cite{Godfrey:1985xj,Isgur:1989vq}. In general, these models
predict more states than those reported by the PDG, leading to the
idea of {\it missing} states in the QCD spectrum. Moreover, apart from
the conventional mesons and baryons, it has been conjectured the
possible existence of exotic states, i.e. those with exotic quantum
numbers like tetraquarks, pentaquarks or hybrid states, all of them
forming color neutral states. Our recent studies on the Polyakov loop
and the corresponding Entropy shift due to a heavy quark, suggest that
conventional hadrons from the PDG or RQM are not enough to saturate
the sum rules, and there are in the spectrum: i) conventional missing
states~($[Q \bar q]$ and $[Qqq]$), and ii) hybrid states ($[Q\bar q
  g]$ and $[Qqqg]$)~\cite{Megias:2012kb,Arriola:2014bfa,Megias:2016onb}.

While the HRG approach has been very successful in the description of
the equation of state (EoS) of the confined phase of QCD, leading to a
good agreement with the lattice data for $T \lesssim 0.8 T_c$~(see
e.g. Refs.~\cite{Megias:2012hk,Arriola:2014bfa} and cited references),
above this regime the HRG assumptions are invalid and the approach breaks
down. In fact, the partition function becomes divergent at some finite
value of the temperature (the so-called Hagedorn temperature) after
summation over a spectrum with an exponential growth of the density of
states, i.e.
\begin{equation}
Z_{\HRG}  = \Tr \, e^{-H_{\HRG}/T}  \stackrel[T \to T_H^-]{\longrightarrow}{} \frac{A}{T_H-T}  \,, \qquad T_H \approx 150 \MeV \,. \label{eq:ZHRG}
\end{equation}
The EoS is sensitive to the spectrum of QCD as a whole, and it is
interesting to study other thermal observables which allow to
distinguish between different channels of the spectrum. In this
contribution we will study the fluctuations and correlations of
conserved charges, and present them as a tool to check the validity of
the HRG approach, as well as to help in the characterization of exotic
and missing states in several sectors of the spectrum of QCD.

\section{Fluctuations of Conserved Charges in a Thermal Medium}
\label{sec:Fluctuations}

Fluctuations of conserved charges $([Q_a,H]=0)$ are a way of selecting
quantum numbers~\cite{Asakawa:2015ybt}. In the ({\it uds}) flavor
sector the only conserved charges are the number of $u$, $d$ and $s$
quarks or equivalently the electric charge~$Q$, the baryon number~$B$,
and the strangeness~$S$. We will study in this section the HRG
realization of the thermal fluctuations of these quantities.


While in the hot vacuum (with no chemical potentials) the thermal
expectation values of conserved charges are vanishing, i.e. $\langle
Q_a \rangle_T = 0$ where $Q_a \in \{ Q, B, S \}$, they present statistical
fluctuations, characterized by susceptibilities and defined
as~\cite{Bazavov:2012jq,Bellwied:2015lba}
\begin{equation}
\chi_{ab}(T) \equiv \frac{1}{V T^3}  \langle \Delta Q_a \Delta Q_b \rangle_T \,, \qquad \Delta Q_a = Q_a - \langle Q_a \rangle_T \,.
\end{equation}
These quantities can also be computed as $  \chi_{ab}(T) \sim \frac{\partial^2 \Omega}{\partial\mu_a \partial\mu_b}  \big|_{\mu_{a,b} = 0}$
from the grand-canonical
partition function, given by
\begin{equation}
Z = \textrm{Tr} \exp\bigg[ - \Big( H - \sum_a \mu_a Q_a \Big)/T \bigg] \,, \qquad \Omega = -T \log Z \,.
\end{equation}


Within the HRG model, the charges are carried by various species of hadrons, $Q_a = \sum_i q_a^i N_i$, where $q_a^i \in \{ Q_i, B_i , S_i\}$ is the charge of the $i$th-hadron corresponding to symmetry $a$, and $N_i$ is the number of hadrons of type $i$. Hence, the susceptibilities are computed as~\cite{RuizArriola:2016qpb,Megias:2017qil}
\begin{equation}
\chi_{ab}(T) = \frac{1}{VT^3} \sum_{i,j \in {\rm Hadrons}} \!\! q_a^i q_b^j \langle \Delta N_i \Delta N_j \rangle_T \,, \quad a,b \in \{ Q, B, S \} \,.
\end{equation}
The averaged number of hadrons is~$\langle N_i \rangle_T = V \int
\frac{d^3k}{(2\pi)^3} \frac{g_i}{e^{E_{k,i}/T} - \xi_i}$, with
$E_{k,i} = (k^2 + M_i^2)^{1/2}$, $g_i$ is the degeneracy and $\xi =
\pm 1$ for bosons/fermions. Since the different species are
uncorrelated, then $\langle \Delta n_\alpha \Delta n_\beta \rangle_T =
\delta_{\alpha\beta} \langle n_\alpha \rangle_T (1 + \xi_\alpha
\langle n_\alpha \rangle_T)$ for the occupation numbers, where
$\alpha$ and $\beta$ stand for any complete set of quantum
numbers. Since $\langle n_\alpha \rangle_T \ll 1$, then~$\langle
\Delta Q_a \Delta Q_b \rangle_T \approx \sum_{i\in {\rm Hadrons}}
q_a^i q_b^i \langle N_i \rangle_T$. The remarkable good agreement of
the EoS found between PDG and RQM~\cite{Megias:2012hk,Arriola:2014bfa}
compared with lattice QCD, while still reasonable for $T \lesssim 150$
MeV, gets a bit spoiled in terms of fluctuations leading to the
conclusion that fluctuations may serve as a diagnostic tool to study
missing states in the spectrum. For instance, the $BB$ susceptibility
suggests that the RQM has too many baryonic states but not too many
charged states~\cite{RuizArriola:2016qpb,Megias:2017qil}. Eventually
at the Hagedorn temperature, $T_H$, the HRG susceptibilities diverge,
$\chi_{ab}|_{\rm HRG} \sim 1/(T_H - T)$, overcoming the quark model
values expected in the large temperature limit; for $uds$ quarks one
should get $\chi_{QQ} \to \sum e_q^2 \equiv 2/3 $, $\chi_{BB} \to
1/N_c $ and $\chi_{SS} \to 1 $.

\section{Local Correlations in a Thermal Medium}

After the overall success of the HRG for the susceptibilities, it is
tempting to extend the analysis at the local level. We are thus
interested in the computation of the retarded correlators of conserved
quantities 
\begin{equation}
C^{ab}_{\mu\nu}(x) = \langle j_\mu^a(x) j_\nu^b(0)\rangle  \,.
\end{equation}
The static version of these correlators are related to the susceptibilities
$\chi_{ab}(T)$ through
\begin{equation}
\chi_{ab}(T) = \frac{1}{T^3} \int d^3x \, C_{00}^{ab}(0,\vec{x}) \,.
\end{equation}

In the case of spin $1/2$ particles the vector currents
are defined as~$j^{\mu}_a(x) = \overline{\Psi}(x) \gamma^\mu {\widehat
  Q}_a \Psi(x)$ with ${\widehat Q}_a$ a matrix that specifies the
charge. In this case the static part behaves at small distances
as~$\langle j_0^a(\vec{x}) j_0^b(0)\rangle \sim r^{-6}$ with $r =
|\vec{x}|$. We can extend this result to finite temperature by using
the Poisson's summation formula~\cite{Megias:2004hj}. From a
comparison of the zero and finite temperature correlators, one finds
that the thermal corrections in the static correlator at small
distances start at ${\cal O}(r^{-2})$ in the case of spin $1/2$
particles.

Within the HRG model, we need to evaluate
\begin{eqnarray}
&&C^{ab}_{\mu\nu}(x) = \sum_{M \in {\rm Mesons}} \!\!\! \frac{1}{2} g_M q^a_M q^b_M C_{\mu\nu}^{J_M}(x) + \!\!\!\!\!\! \sum_{B \in {\rm Baryons} > 0} \!\!\!\!\!\! g_B q^a_B q^b_B C_{\mu\nu}^{J_B}(x) \,,  \label{eq:corr_HRGM}
\end{eqnarray}
where $C^J_{\mu\nu}(x) \equiv \langle j_\mu(x) j_\nu(0) \rangle_J$ are
the correlators of free particles of spin $J$. The summations in $M$
and $B$ run over the spin multiplets of mesons and baryons, each of
them with degeneracy $g_M = (2J_M+1)$ and $g_B = (2J_B+1)$
respectively. Baryons and antibaryons contribute to the correlators in
the same amount, so that we have considered in $\sum_B$ a summation
over baryons only, and multiplied it by a factor $2$. It is possible
to extend the analysis to particles of any spin by using the
Bargmann-Wigner formalism~\cite{Bargmann:1948ck} and to obtain the
correlators in Euclidean space (details will be provided
elsewhere~\cite{Megias2018}). The small distance behavior of the
static correlators, either at zero or finite temperature, reads (for
$J>0$ and up to a factor)
\begin{equation}
C^{ab\, J}_{00}(r) \stackrel[r \to 0]{\sim}{} \delta_{ab}
\frac{m^2}{r^4} \frac{1}{(mr)^{4J}}
\,. \label{eq:correJ_smalldistance}
\end{equation}

\begin{figure}[h]
\begin{center}
\epsfig{figure=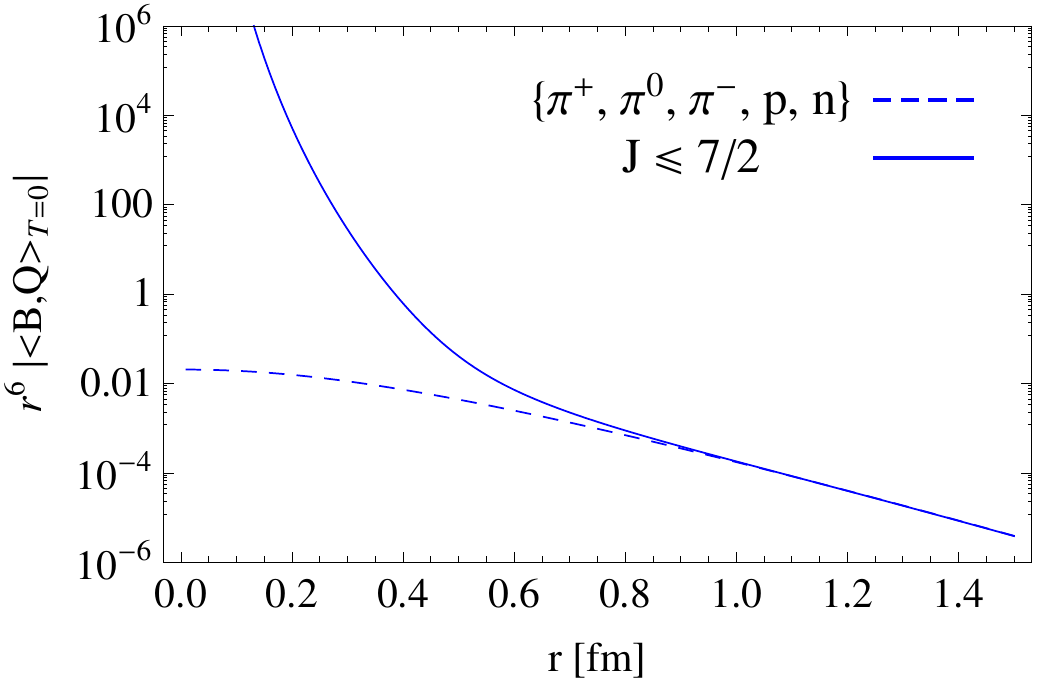,width=56mm} \hspace{1.0cm}
\epsfig{figure=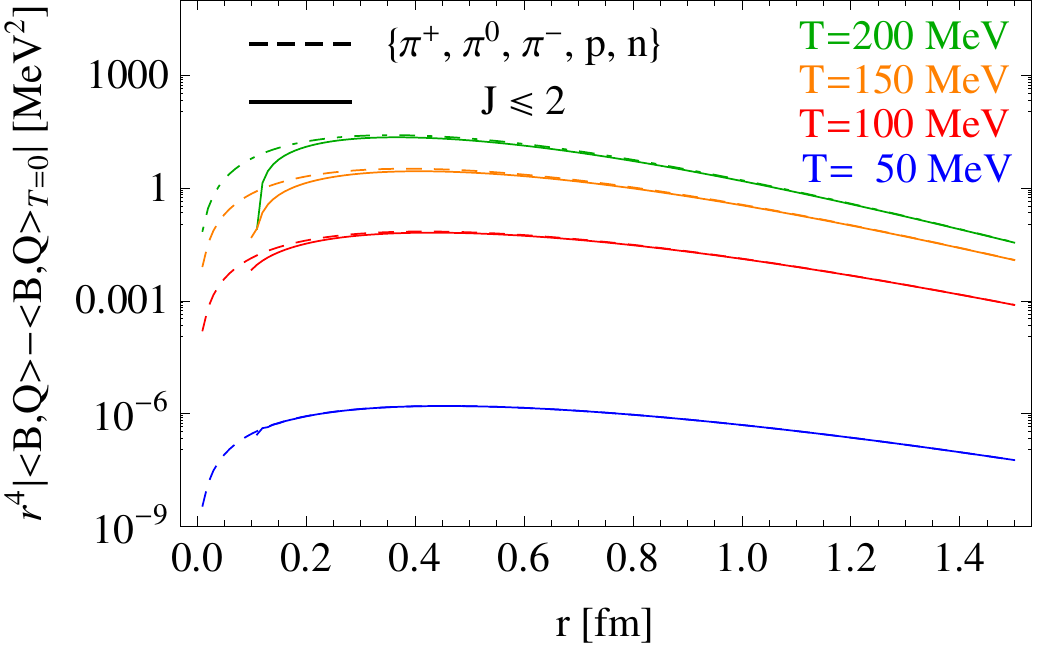,width=56mm}
\end{center}
\caption{Static correlator $C_{00}(0,\vec{x}) \equiv \langle B(\vec{x}) Q(0) \rangle$ at zero temperature (left panel), and finite temperature (right panel), including pions and nucleons (dashed lines), and hadrons with $J \le 7/2$ and $J \le 2$ from the RQM spectrum~\cite{Godfrey:1985xj,Isgur:1989vq} (solid lines).}
\label{fig:corr}
\end{figure}

 As an example, the lowest lying states contributing to the summations
 of Eq.~(\ref{eq:corr_HRGM}) are $M \in \{ \pi^+ , \pi^- , \pi^0\}$
 and $B \in \{ p , n\}$.  We display in Fig.~\ref{fig:corr} the result
 of the static correlators at zero and finite temperature in the
 confined phase of QCD, within the HRG model. Notice the important
 growth of the zero temperature correlator at short
 distances. Unfortunately, to the best of our knowledge there are no
 lattice studies for these quantities yet.

\section{Dual Hagedorn distance in the correlators}

We may conjecture the existence of a formal analogy between the static
correlators at zero temperature and the finite temperature susceptibilities. From the asymptotic behaviors of these
two quantities,
\begin{equation}
C_{00}^{ab}(0,\vec{x}) \big|_{T=0} \stackbin[r \to \infty]{}{\sim} e^{-2m r} \quad \textrm{and}  \quad \chi_{ab}(T) \stackbin[T \to 0]{}{\sim} e^{-m/T} \,,
\end{equation}
where $m$ is the mass of the lowest-lying state, one
concludes that they have a similar behavior after considering the
replacement $r \leftrightarrow 1/T$. In particular, the existence of a
limiting temperature for the validity of the hadronic representation
of $\chi_{ab}(T)$, i.e. $T < T_H$, cf. Eq.~(\ref{eq:ZHRG}), might have
its counterpart in the existence of a limiting distance in the
hadronic representation of the correlators $(r > r_H)$. In fact, given
the behavior of the static correlators for particles of spin $J$,
cf. Eq.~(\ref{eq:correJ_smalldistance}), we expect
that after summation over hadrons of higher and higher spin, the
static correlators within the HRG model present a divergence at some
finite value of the distance
\begin{equation}
C_{00}^{\HRG}(r) = \sum_J C_{00}^J(r)  \stackrel[r \to r_H^+]{\longrightarrow}{} \infty \,.
\end{equation}
 When using $T_H \approx 150 \MeV$ one gets $r_H \simeq 1/(2\pi T_H)
 \approx 0.21 \fm$, where the factor $1/(2\pi)$ is standard in finite
 $T$ computations. Notice that this value of $r_H$ is of the same
 order of magnitude than the distance at which the growth in
 Fig.~\ref{fig:corr} (left) become significant, when including hadrons
 with $J \le 7/2$. This would be analogous to the divergence of the
 partition function at the Hagedorn temperature, after summation over
 a spectrum with an exponential growth of the density of states. As in
 the susceptibilities case the divergence appears as a purely hadronic
 feature, but we find this to happen for smaller distances than $0.6 \fm$~\cite{Megias2018} where the hadronic correlation overcomes the quark correlation which should dominate for $r\to 0$.

\section{Conclusions}
\label{sec:conclusions}

At very low temperatures hadrons can be considered as a complete basis
of states in terms of a HRG model. However, close to the deconfinement
crossover of QCD, many hadrons are needed to saturate the sum rules,
so that this regime turns out to be very interesting for the
characterization of missing states in the spectrum. In this work we
have argued that fluctuations and correlations of conserved
charges can be used to study the existence of missing and exotic
states in three different sectors: i) electric charge, ii) baryon
number, and iii) strangeness. Finally, we have conjectured a duality
between zero temperature correlators and finite temperature
susceptibilities. This duality leads to the appearance of a dual
Hagedorn distance in the correlators. It would be desirable to
confront these results for the correlators with future results on the
lattice.



\end{document}